\documentclass[a4paper,11pt]{amsart}
\usepackage{amssymb,amsthm,palatino}
\usepackage{graphicx}
\newtheorem{theorem}{Theorem}
\newtheorem{corollary}[theorem]{Corollary}
\newtheorem{lemma}[theorem]{Lemma}

\theoremstyle{definition}

\newtheorem{remark}[theorem]{Remark}
\newtheorem{example}[theorem]{Example}

\newcommand{\K}{\mathbb{K}}
\newcommand{\F}{\mathbb{F}}
\begin{document}

\title{Managing key multicasting 
  through orthogonal systems  }

\thanks{The Research was supported in part by the Swiss National Science
Foundation under grant No. 149716. First author is partially supported
by Spanish Ministry of Science and Innovation (TIN2008-01117), and
Junta de Andaluc\'{\i}a (P08-TIC-3518).  Second author is partially
supported by by Spanish Ministry of Science and Innovation
(TEC2009-13763-C02-02) and Junta de Andaluc\'{\i}a (FQM0211). }

\author{Jos\'e Antonio  Alvarez-Bermejo}
\address[Jos\'e Antonio  Alvarez-Bermejo]{Departamento de \'Algebra y An\'alisis Matem\'atico
Universidad de Almer\'ia
04120 Almer\'ia, Spain}

\author{Juan Antonio Lopez-Ramos}
\address[Juan Antonio Lopez-Ramos]{Departamento de \'Algebra y An\'alisis Matem\'atico
Universidad de Almer\'ia
04120 Almer\'ia, Spain}
\urladdr{www.ual.es/$\sim$jlopez}

\author{Joachim Rosenthal}
\address[Joachim Rosenthal]{Mathematics Institute, University of Zurich, 
CH-8057 Zurich,  Switzerland}
\urladdr{www.math.uzh.ch/aa} 

\author{Davide Schipani}
\address[Davide Schipani]{Mathematics Institute, University of Zurich, 
CH-8057 Zurich,  Switzerland}
\urladdr{www.math.uzh.ch/aa}

\date{December 31, 2014}

\thispagestyle{empty}

\begin{abstract}
  In this paper we propose a new protocol to manage multicast key
  distribution. The protocol is based on the use of orthogonal systems
  in vector spaces.  The main advantage in comparison to other existing
  multicast key management protocols is that the length and the number
  of the messages which have to be sent are considerably smaller. This
  makes the protocol especially attractive when the number of legitimate
  receivers is large.
\end{abstract}
\maketitle

\vspace{2mm}



\section{Introduction}

Traditional security measures are mainly applicable to a unicast
environment, i.e. communications take place between two single
parties. For instance, data confidentiality, one of the most important
features in network security, can be offered in this environment by
means of a pair of keys. However there exist many different situations
where the usual secure unicast protocols cannot be used, mainly due to
the nature of the information to be transmitted. This usually happens
when trying to deliver data from a sender to multiple receivers,
especially when a huge amount of data needs to be delivered very
quickly. One of the most efficient ways to do this is the so-called
multicast. In a multicast protocol a certain group of people receives
the information and this group is usually highly dynamic. In a typical
situation users join and leave the group constantly (\cite{survey}).

There are a number of exciting multimedia applications that make good
use of multicast capability, such as stock quote services,
video-conferencing, pay-per-view TV, Internet radio, and so on.
Many of the aforementioned multicast applications require security in
data transmission, i.e., data can only be accessed or exchanged among
an exclusive group of users. In the Pay-TV system, for example, the
service providers employ Conditional Access System (CAS) to avoid
unauthorized accessing of their video/audio streams, and only allow
access to services based on payment.

The natural approach to establish secure multicast communications is
to agree on one or several symmetric encryption keys in order to
encrypt messages. However, the key, or keys, must be renewed
periodically to prevent outer or inner attacks.  Depending on how key
distribution and management are carried out, secure multicast schemes
are divided into centralized and distributed schemes.  Centralized
schemes depend directly on a single entity to distribute every
cryptographic key. A typical scenario is an IPTV or P2PTV platform, in
which clients receive a TV signal from a Content Server via Internet.
Distributed schemes are able to manage higher number of audiences but,
on the other hand, key management involves other problems that make
them more complex (\cite{survey}). A big issue concerns security: in a
centralized system there is just one server to secure, while in the
distributed one security efforts have to be multiplied. Our aim in
this paper is to introduce a novel protocol applicable for centralized
multicast that is shown to be secure, efficient and scalable.

In the following lines we recall some centralized schemes for key
management, although the reader can find a recent survey on secure
multicast in \cite{scalable}. A very well-known protocol is {\it
  Hierarchical Tree Approach} (HTA) \cite{RFC:2627}.
It uses a logical tree arrangement of the users in order to facilitate
key distribution. The benefit of this idea is that the storage
requirement for each client and the number of transmissions required
for key renewal are both logarithmic in the number of members. Other
key tree approaches and extensions are LKH \cite{332988}, LKH++
\cite{Pietro02efficientand}, OFT \cite{OFT2003} or ELK \cite{ELK}.

In \cite{72912} the so-called {\it Secure Lock} protocol is
introduced.  The authors approach the problem in a computational
manner and make use of the Chinese Remainder Theorem instead of a tree
arrangement.  Users are distributed into groups, that in the case of
PayTV could be represented by those subscribers with the same
Pay-Per-Channels (PPC) or Pay-Per-View (PPV) options. The PPC and PPV
programs should be encrypted previously to their distribution and
there is only a content server and a key server (that might be
different or not).  Its main drawback is the large computational cost
required at the key server side on each rekeying operation: the length
of the rekeying messages and the computing time needed becomes quickly
problematic as the number of members in one of the PPC or PPV groups
grows \cite{Naval98techniquesand}.

In \cite{1039707}, a divide-and-conquer extension of the Secure Lock
is proposed. It combines the Hierarchical Tree Approach and the Secure
Lock: members are arranged in a HTA fashion, but the Secure Lock is
used to refresh keys on each tree level. Therefore, the number of
computations required by the Secure Lock is reduced.

Another computational approach with the same distribution by groups of
users and a unique key server is introduced in \cite{cas} with the
particular application on Pay-TV but extendable to any other secure
multicast application. 
The idea is to use polynomials over a finite field interpolating
hashes of secret values belonging to the authorized users. The main
drawbacks are that the hash function must be renewed with any rekeying
operation, due to security concerns, and the large size of the
polynomials involved. The length of the messages grows linearly with
the number of users in every group, so that if this number is huge,
users might be forced to be distributed into subgroups, e.g., groups
of users are established inside every PPC or PPV group.

The distribution by groups is in fact often beneficial and is used by
most key managing protocols. A first benefit is the parallelization of
the process which speeds up the rekeying operations. Secondly a
compromised key in one of the groups does not affect the others.  Last
but not least, 
in most applications of secure multicast the group distribution is
connected with the scalability of the system, i.e., the efficiency of
the communication protocols concerning the rekeying process, with
particular reference to leave and join operations. Groups are usually
highly dynamic and the joining or the leaving of users implies a
rekeying operation, and thus key refreshment due to this fact in one
group does not affect the others.

More recently in \cite{euclid} the authors introduce another solution
with the same philosophy of Secure Lock and of that introduced in
\cite{cas} and based on the Extended Euclidean Algorithm. Throughout
this paper we will refer to this protocol as {\it Euclides}. The
server distributes a secret via the inverse of an integer modulo a
product of coprime secret numbers, each one of them belonging to an
authorized user. The authors show \cite{euclid} that a former user
could try a factorization attack, which forces to consider prime
numbers of an adequate size. This implies a division by groups of the
audience, in the case of PayTV a subdivision of every PPC or PPV
group, since the length of the rekeying messages could become
unaffordable as in the other computational
approaches. 

In this paper we introduce a new protocol for key managing in
centralized multicast. We are assuming a scenario, fairly general and
suitable for many applications, especially for multimedia distribution
purposes, with a Key Server and a set of members (other hosts) that
either send or receive multicast messages. Any multicast topology can
be used underneath. All setup tasks are carried out by the Key Server.
Data communications are then either one-to-many or many-to-many, and
consist of encrypted contents and/or rekeying messages, which are
created by the Server (or the two servers, Content Server and Key
Server).  Members can enter and leave the system at any time. The key
must be refreshed upon member arrival or departure to achieve perfect
backward and forward secrecy, respectively. However this might depend
on the application, since there exist cases, such as some audio and/or
video streaming distributions, where backward secrecy is not an issue,
as contents can be out of date.

The protocol we are introducing presents some nice features that make
it competitive, e.g. it requires just a single message per group, of
affordable length, for every rekeying operation, the operations at the
Key Server and the Client sides require low computational cost and the
key storage requirements are minimal.

The main idea behind the protocol is the use of orthogonal systems in
vector spaces. Exploiting orthogonality comes probably as a natural
tool in multicast applications, as this appears also e.g. in CDMA and
\cite{wang}. How orthogonality is exploited here appears though to be
new, and brings with it several advantages. In particular the scalars
will play an important role in order to have fast rekey and reset
operations and avoid involving large vectors to be replaced or
generated. Moreover this structure will make the protocol not only
agile and flexible, but also more robust and secure against all
conceivable attacks, as will be shown later.

In the next Section we describe the new protocol, analyse its
security, and compare it with other existing and aforementioned
protocols. 
Section III demonstrates an efficient implementation of the
protocol. 

\section{The proposed protocol}

Let the potential users be denoted with the integers $1,\ldots,n$ and
assume that they all belong to the same group.
\begin{enumerate}

\item Initialization step:

  Let $\K$ be a field and $V$ be a $\K$ vector space of dimension
  $m\geq n$ (see also next subsection for the choice of $m$).  Let
  $<,>$ be a bilinear form which we assume to be nondegenerate and
  symmetric.  Let $B=\{ \mathbf{e_1}, \dots ,\mathbf{e_n} \}$ be a set
  of $n$ mutually orthogonal vectors in $V$ having the property that
  $<\mathbf{e_i},\mathbf{e_i} >\neq 0$ for $i=1,\ldots,n$. Note that these two requirements, namely the mutual orthogonality and the non self-orthogonality, also imply the linear independence of the vectors. For security reasons, as we will show later, the vectors should not be part of the
  canonical basis or anyway the basis used to represent vectors.  We
  select a family $\{ x_i \}_{i=1}^n$ of random nonzero scalars in
  $\K$.  Note that $B'=\{ x_1 \mathbf{e_1}, \dots , x_n \mathbf{e_n}
  \}$ spans the same subspace as $B$. These two sets are kept secret
  by the server and each user $i$ is assigned the vector
  $\mathbf{v_i}=x_i\mathbf{e_i} $. By our assumptions we know that
  $<\mathbf{v_i} ,\mathbf{v_i} >\neq 0$ for $i=1,\ldots,n$. Then we
  will consider two subsets in $B'$ at a determined point in time in
  the communications. On one hand $B'_1$ will be formed by those
  vectors in $B'$ that are assigned to some user and $B'_2$ will be
  the set of vectors in $B'$ that do not correspond to any user. We
  also consider two subsets in $B_2'$, $B'_{2,1}$ and $B'_{2,2}$, that
  contain those vectors that were not previously used and those that
  are a multiple of a vector corresponding to a former no more
  legitimate user, respectively.

\item Sending the information:

  Suppose that we want to distribute the secret $s\in \K$. Then we
  compute 
  and multicast (broadcast) $\mathbf{c} =s(\sum_{\mathbf{v_j}\in
    B'_1}\mathbf{v_j}+y\sum_{\mathbf{v'_j}\in B'_2}\mathbf{v'_j}
  )$ 
  for some random $y\in \K$ different for each new secret.
\item Recovering the information:

  Each user computes $h=<\mathbf{c} ,\mathbf{v_i} >=s<\mathbf{v_i}
  ,\mathbf{v_i} >$.

  The secret $s$ is then recovered by computing $s=h<\mathbf{v_i}
  ,\mathbf{v_i} >^{-1}$.

\item Key refreshment:

  \begin{enumerate}

  \item Join:

    If user $j$ joins, then she is assigned one of the vectors in
    $B'_{2,1}$, say $x_j\mathbf{e_j} $. The server selects a new
    secret $s'\in \K$ and multicasts as above $\mathbf{c'}
    =s'(\sum_{\mathbf{v_j}\in
      B'_1}\mathbf{v_j}+y'\sum_{\mathbf{v'_j}\in B'_2}\mathbf{v'_j} )$
    for some random $y'\in \K$.

  \item Leave:

    If user $j$ leaves, then her vector $\mathbf{v_j} =x_j\mathbf{e_j}
    $ is deleted from $B'_1$ and the vector
    $\mathbf{v'_j}=x'_j\mathbf{e_j}$ is included into $B'_{2,2}$ where
    $x'_j$ ($\neq x_j)$ is selected at random in $\K$. To distribute a
    new secret $s'$, we do similarly as after a join
    operation. 

  \end{enumerate}

\end{enumerate}

\begin{figure}
  \begin{center}

    \includegraphics[width=9cm]{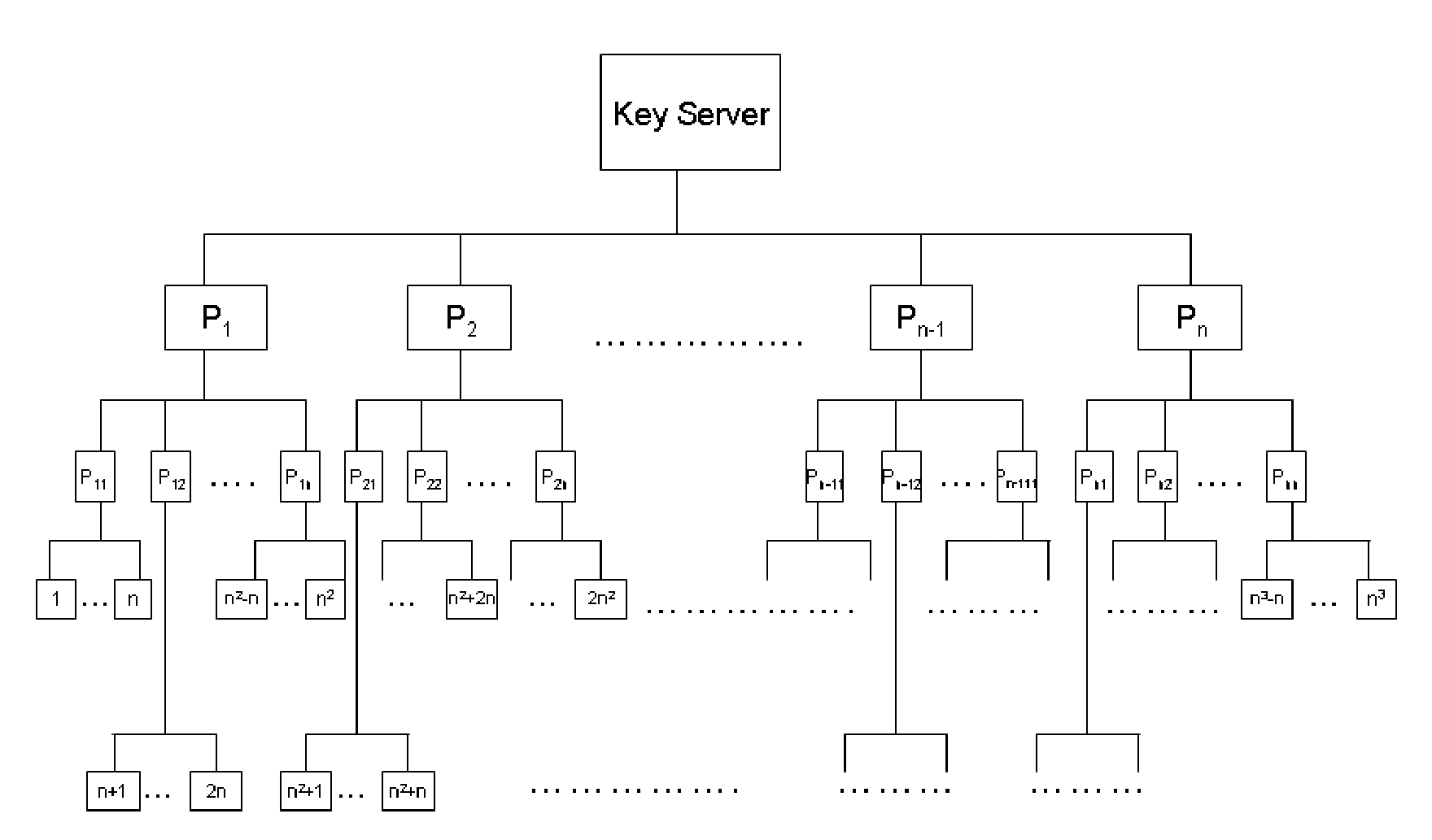}
    \caption{Hierarchical Tree}
    \label{fig:tree}
  \end{center}
\end{figure}

We remark that, if we are managing with $\ell$ groups, then we can use
$\ell$ different orthogonal bases.  A particular interesting case of
managing groups is shown in Figure \ref{fig:tree}. In this case one
can profit from a tree-like distribution as in the HTA and the Secure
Lock + HTA approaches (\cite{RFC:2627} and \cite{1039707}
respectively) using a divide and conquer
strategy. 
The main advantage is that we can use smaller vector spaces, so that
we can serve a much bigger number of users without delaying in
rekeying operations. 
Let us consider for example a hierarchical tree with a depth of 4,
i.e., the number of levels below root is 3, and a degree of $n$, i.e.,
the number of children below each parent node is $n$ (see Figure 1).
In this situation, we have $n^3$ users who are located at the leaves
of the tree. Intermediate nodes store group keys in the form of
vectors known to the correspondent descendants. For example, users $1$
to $n^2$ share a common vector stored at $P_1$, users $1$ to $n$ share
another common vector stored at $P_{1,1}$ and so on, so that user $1$
knows her private vector plus the vectors $P_{1}$ and $P_{1,1}$. Then,
when a rekeying message is to be sent, this will be made using the
orthogonal system given by $\{ P_1, \dots ,P_n \}$ as described above,
and any authorized user will be able to retrieve the key.

Now, let us assume without loss of generality that user $1$ wants to
leave. It can be easily observed that we need the following messages
to refresh the key and preserve forward secrecy.  After determining a
new vector for position $1$, the server uses the new basis, including
the private vectors of users $2$ to $n$, to send them a scalar (by
posting an encrypted version, as described in the protocol). Users $2$
to $n$ use this scalar to renew $P_{1,1}$, i.e. they keep the same
vector but substitute the scalar associated to it. Analogously the
server uses the new basis for $P_{1,1}$ to $P_{1,n}$, to send the
users $1$ to $n^2$ another scalar. These use this scalar to renew
$P_{1}$. At last the server can send a new key for all users using the
new orthogonal system for $\{ P_1, \dots ,P_n \}$.

Notice that in order to send a scalar of $k$-bits length we need a
message of $nk$ bits length. If we deal with a tree distribution as
above where $n=100$, then three rather short messages will give us the
possibility of handling audiences of up to 1 million users, and the
computing time to generate the rekeying messages will not depend on
the number of users.





In fact, our protocol is natural from the point of view of building
the tree. When designing the tree distribution we have to fix the
number of descendants of each node, that will give us the required
dimension for our vector space.  And the tree distribution allows to
have flexibility on the number of users: if a bigger structure has to
be considered in order to deal with more users, then intermediate
nodes can be easily inserted.



\subsection{Security}

Let us suppose that $\K$ is a finite field, which is the usual setting for application. As a first step we will show that, by choosing $m$ appropriately, we
can be sure that there are sufficiently many $n$-tuples of mutually
orthogonal vectors in $V$, so that a brute force attack to find $B$ is
not feasible.

As we require also the property $<\mathbf{e_i},\mathbf{e_i}>\neq 0,$ for
$i=1,\ldots,n$, we consider the set 
$\{\mathbf{x}=(x_1,\ldots,x_m)\in V |
<\mathbf{x},\mathbf{x}>=x_1^2+\cdots+x_m^2=0\}$. This set forms a hypersurface $H$ of
dimension $m-1$ and degree 2.

\begin{theorem}
  Let 
$V:=\F_q^m$ and
  $m > 2n$. Then there exists at least
$$
(1+o(1))\frac{q^{\frac{3}{2}n^2}}{n!}
$$
$n$-tuples of mutually orthogonal vectors
$(\mathbf{e_1},\ldots,\mathbf{e_n})$ in $V$ having the property that
$<\mathbf{e_i},\mathbf{e_i}>\neq 0, i=1,\ldots,n.$ For characteristic bigger than $2$ we may require $(1+\frac{5\cdot 2^{13/3}}{q})q^{m-1}\ll q^{m}$.
\end{theorem}

\noindent{\it Proof.}  
We divide the proof in different scenarios, as the estimate can be
made more precise depending on the particular setting involved.

In characteristic $2$ (which is probably the most interesting case for
cryptographic applications), $x_1^2+\cdots+x_m^2=(x_1+\cdots+x_m)^2$
so that the cardinality of the hypersurface $|H|=q^{m-1}$.  Now
Iosevich and Senger~\cite{iosevich,vinh} derived a lower bound on the
number of $n$-tuples of mutually orthogonal vectors in a subset
$X\subset V$ in situations where a lower bound on the cardinality of
$X$ is known. Applying this result to our situation with
$X:=V\setminus H$ one derives the thesis.

In characteristic greater than $2$, we can exploit estimates on the
number of points in hypersurfaces, namely the Lange-Weil bound and
connected results (e.g.\cite{cafure} or\cite{la14p} where probably the 
best general bounds can be found). If $H$ is absolutely
irreducible, then $|H|=(1+C)q^{m-1}$, where $|C|$ can be estimated,
independently of any regularity conditions, as less than $\frac{5\cdot
  2^{13/3}}{q}$ (\cite[Theorem 5.2]{cafure}); here we may require $q$
to be large enough to make $|H|$ neglectable with respect to $q^m$. If
$H$ is not absolutely irreducible, then $|H|\leq q^{m-1}$ (\cite[Lemma
2.3]{cafure}). In any case we can apply again the same argument as in
characteristic $2$ and derive the thesis. $\Box$

\medskip 

The condition $m > 2n$ might also be convenient in order to avoid any
collusion attack, that is to avoid that a big group of, say, $k$ users
share their private vectors with each other, trying to retrieve
information belonging to other authorized users. Since $m-k>2(n-k)$,
the inequality above guarantees that there would be anyway more than
$$
(1+o(1))\frac{q^{\frac{3}{2}(n-k)^2}}{(n-k)!}
$$
$(n-k)$-tuples of mutually orthogonal vectors in the remaining unknown
vector space.


Let us assume now that the set $B$ is known, instead of being kept secret. Since $B$
is a linearly independent set, one can compute readily the unique
coefficients $z_1,\ldots,z_n$ such that
$$
\mathbf{c} =z_1 \mathbf{e_1} + \dots +z_n \mathbf{e_n} .
$$
An authorized user knowing the vector $\mathbf{v_j} =x_j\mathbf{e_j} $
and having computed $z_j\mathbf{e_j} $ readily computes $x_j$ and $s$
from $z_j=sx_j$. With this all the private numbers $x_i$,
$i=1,\ldots,n$ can be readily computed by this user. Such a user would
have the chance to use this later in her own interest. As it is often
the case, inner attacks are more dangerous than outer ones.

The security is clearly compromised not only if the set $B$ is made
public, but also if just one vector of $B'$ becomes known to
unauthorized users: in fact getting $s$ involves knowing at least one
vector in the set $B'$ used to compute $\mathbf{c} $. We can think at
different ways for an attacker to get such a vector. Let us assume in
the following without any loss of generality that the set $B'_{2,2}$
is formed by just one vector corresponding to a single former user.




First, the former user can try to get the new $s'$ using her vector,
say $\mathbf{v_i} $. If she multiplies $<\mathbf{c'} ,\mathbf{v_i} >$,
then she gets
$$
<\mathbf{c'} ,\mathbf{v_i} >=<s'(x_1\mathbf{e_1} +\dots
+yx'_i\mathbf{e_i} +\dots +x_r\mathbf{e_r} ),x_i\mathbf{ e_i}>
=s'yx_ix'_i<\mathbf{e_i} ,\mathbf{e_i} >.
$$
\noindent for some random $y$. But now she would have to know the
vector $\mathbf{e_i} $ (or equivalently $x_i$) and the value $yx'_i$
to get the new secret
$s'$. 
Also, knowing $s'$ does not reveal anything on $x_i$, nor
$\mathbf{e_i}$.


Another option consists in trying to derive some information from the
difference between two different rekeying messages $\mathbf{c} $ and
$\mathbf{c'} $. But
$$\mathbf{c} -\mathbf{c'} =(s-s')x_1\mathbf{e_1} +\dots +(sx_i-s'yx'_i)
\mathbf{e_i} +\dots +(s-s')x_r\mathbf{e_r} $$ Then $<\mathbf{c}
-\mathbf{c'} ,\mathbf{v_i} >=(sx_ix_i-s'yx'_ix_i)<\mathbf{e_i}
,\mathbf{e_i} >$ and, as before, no information can be deduced about
$s'$.


Let us assume now that the attacker has additional means, for example
suppose a valid user shares $s$ with a former user. 

First note that 
  forward secrecy is not violated under a known plain text attack.
Indeed
suppose the former user knows $\mathbf{v_i} =
x_i\mathbf{e_i}$ for some $i$ and let $\mathbf{v'_i}=x'_i\mathbf{e_i}$
be the unique element of $B'_{2,2}$. With the rekeying procedure
$\mathbf{c} = s(x_1\mathbf{e_1}+ \dots + yx'_i\mathbf{e_i} + \dots +
x_r\mathbf{e_r})$ for some random $y$ is multicasted. Then
$<\mathbf{v_i},\mathbf{c}>=sx_iyx'_i<\mathbf{e_i}, \mathbf{e_i}>$.  If
somehow this former user has access to the corresponding decrypted
message, $s$, then she will be able to get $x_iyx'_i<\mathbf{e_i},
\mathbf{e_i}>$ by multiplying by $s^{-1}$, but this cannot be used for
the following multicasted messages since $y$ is chosen randomly with
every rekeying stage. 


If $y$ was not recomputed with every rekeying message, then given a
new $\mathbf{c'} = s'(x_1\mathbf{e_1}+ \dots + yx'_i\mathbf{e_i} +
\dots + x_r\mathbf{e_r})$ for the same $y$ used for $\mathbf{c}$, then
$s'$ would result from computing
$<\mathbf{c'},\mathbf{v_i}>(x_iyx'_i<\mathbf{e_i},
\mathbf{e_i}>)^{-1}$. 

  Forward secrecy is also not violated by a chosen plain text attack. Indeed
assume that an attacker has access to an
algorithm that provides the corresponding $\mathbf{c}$ for any given
message $s$. This is like a valid user that has access to many such
pairs. Then even in this case, as a valid user she could compute for
all pairs of ciphertexts $<\mathbf{c} -\mathbf{c'} ,\mathbf{v_i}
>=(s-s')x_i^2<\mathbf{e_i} ,\mathbf{e_i} >$, but no information on
$\mathbf{ e_i}$ would be leaked. 


Similar arguments apply for new users concerning backward secrecy.


Finally we can imagine an attack based on the collection of many
subsequent pieces of information, in a cipher text-only attack
scenario. We show below that this is feasible when $B$, against our
hypothesis, is the canonical basis used to represent vectors of the
vector space $V$.
Namely, anybody observing the information flow could get $n$ linearly
independent key refreshments $\mathbf{c_1}, \dots ,\mathbf{c_n} $.
Note that this is the case whenever a user $i$ leaves and in that
case, the set $B'=\{ x_1\mathbf{e_1} , \dots ,x_i\mathbf{e_i} , \dots
,x_n\mathbf{e_n} \}$ would change to $B''=\{ x_1\mathbf{e_1} , \dots
,yx'_i\mathbf{e_i} , \dots ,x_n\mathbf{e_n} \}$. Now, suppose without
loss of generality that $n=m$; if the server sends $(s_ix_1,\dots
,s_ix_n)$ as a rekeying message $\mathbf{c_i} =(c_{i,1},\dots
,c_{i,n})$, then we would consider the matrix

$$M=\left( \begin{array}{ccc} c_{1,1} & \cdots & c_{n,1} \\ \vdots &
    \vdots & \vdots \\ c_{1,n} & \cdots & c_{n,n}
  \end{array} \right)$$

\noindent where each column $(c_{i,1},\dots ,c_{i,n})$ represents the
coordinates of the refreshment $\mathbf{c_i} $ with respect to $B$ (as
$c_{i,j}=s_ix_j$ for $i,j=1,\dots ,n$); then $M$ represents the change
of basis from the basis $C=\{ \mathbf{c_1} , \dots ,\mathbf{c_n} \}$
to $B$. The inverse of $M$ will reveal then $B$ in terms of the basis
$C$. And knowing a pair $(s,\mathbf{c} )$ would compromise all the
secrets (of the other users) used to get this pair $(s,\mathbf{c} )$,
as noted at the beginning of this subsection.

Therefore it is convenient, as pointed out before, that $B$ is chosen
not to be the canonical
basis, 
so that what is sent by the server is not plainly $(s_ix_1,\dots
,s_ix_n)$, but its representation in another basis.

Let us illustrate this with the following easy example:

\medskip

\begin{example}
  Let $B= \{ (1,1,1),(1,-2,1),(-1,0,1) \}$ be an orthogonal basis of
  the euclidean vector space $\mathbb{R}^3$ (with the usual scalar
  product $<,>$) and assume $x_1=2,\ x_2=3, \ x_3=5$. Then $B'=\{
  (2,2,2), (3,-6,3),(-5,0,5) \}$.

  If we want to rekey with $s=4$, then we have to multicast
$$
\mathbf{c_1} =4(2,2,2)+4(3,-6,3)+4(-5,0,5)=(0,-16,40).
$$
User 1 can recover $s$ by calculating
$$
h=<(0,-16,40),(2,2,2)>=48,
$$
and then
$$
s=h<(2,2,2),(2,2,2)>^{-1}=4.
$$ 
Users 2 and 3 act similarly.

Now suppose that user 2 leaves and $x_2$ is changed to $x_2=2$. $B'_{2,2}$, that was previously empty, contains now the vector $(2,-4,2)$.  Then
the rekeying message for $s=3$ and considering $y=2$ is $\mathbf{c_2}
=3(2,2,2)+3\cdot 2(2,-4,2)+3(-5,0,5)=(3,-18,33)$. 
Finally, suppose that user 1 leaves, $x_1$ becomes $3$ and the new
secret $s$ is $2$, so that, choosing $y=-1$, $\mathbf{c_3}
=2\cdot (-1)(3,3,3)+2\cdot (-1)(2,-4,2)+2(-5,0,5)=(-8,2,8)$. Now the basis given by $\{
\mathbf{c_1} ,\mathbf{c_2} ,\mathbf{c_3} \}$ does not tell anything
about the basis $B$.
\end{example}

\begin{remark}
  It should be remarked that, if
  we 
  restrict ourselves to work in a subring of the base field that
  admits an algorithm to compute $GCD$s, then $s$ divides the GCD of
  the coordinates of $\mathbf{c} $. 
  Observe, for instance, that in the example above $s$ divides
  $GCD(0,-16,40)$ and after the first rekeying $s=GCD(3,-18,33)$. Thus
  this situation should be avoided for a security issue, so finite fields
  rather than the ring of integers should be
  used. 
\end{remark}

\begin{remark}
  To add additional security and prevent any sort of statistical or
  brute force attacks, it is anyway advisable to perform periodic key
  refreshments, as is commonly done for other
  protocols. 
\end{remark}

\subsection{Comparison with other schemes}

We compare here our new proposal with some of the other key managing
protocols existing in the literature and cited in the introduction.
The main parameters we will focus on are the key storage cost and the
length of the messages.

For additional comparisons, as our protocol behaves comparably to
Euclides in the number of rekeying messages per join or leave, we
refer to \cite{euclid}, in particular Table 1, where Euclides is
compared with previous protocols and other features are also taken
into account.

As for the protocol we are introducing in this paper,
the server has to store one scalar per user, the $x_i$'s, and an
orthogonal system, $B$, for each considered group, each user stores
her vector $\mathbf{v_i} $, while the length of the rekeying messages
is $n\cdot C$, where $C$ is  the bit-length of the elements in a
finite field $I\! \! K$. 


In \cite{euclid}, Euclides was introduced and shown to be already very
competitive with respect to existing protocols, however the present
proposal offers an additional advantage concerning the length of the
messages. In Euclides, in fact, the key storage cost can be of the
same order as here, but the length of the messages could become a
problem unless some key management by groups is used. In fact, by
security requirements every private key held by any user, an integer,
has to be of appropriate length to avoid a factorization attack by a
former user (cf. \cite{euclid}). In this way integers of length 1024
bits onwards should be considered and since the rekeying messages are
of the same order as the product of all these integers, then for large
groups these could be unaffordable.

On the other hand, in this new protocol messages can be considerably
shorter than in Euclides, depending on the number of users in the
group and on the cardinality of the field chosen for the scalars. 

Suppose 
for example that we are dealing with a field of the order 64-bits
length elements and we are using a vector space of dimension 10000.
Then rekeying messages would be shorter than 80Kbytes length, which is
perfectly affordable by any multicast network used for this purpose.
In the case of Euclides, using primes of 64-bits length produces
messages of the same length, i.e. 80Kbytes.
However, 
any user, as it is shown in \cite{euclid}, has access to a multiple of
the product of all the secret keys and so this bit-length of the
primes is not enough for a secure rekeying process since a
factorization attack would succeed very quickly. To avoid this we are
forced to deal with 1024-bits length primes (at least).  This leads to
over 1Mbyte length rekeying messages. Otherwise we have to divide this
audience in at least 12 groups in order to deal with messages of
length comparable to that of the new proposal.

\medskip

In the case of Secure Lock 
each user holds a pair of keys, an integer, $R_i$ and a symmetric key,
$k_i$. The server encrypts the secret using the symmetric key $k_i$ of
every user, obtaining a number for each one of them, $N_i$. Then the
server solves the congruence system $x \equiv_{R_i}N_i$ and multicasts
the solution $U$ of this system.  We observe that, as in the case of
Euclides, the length of the messages is of the same order as the
product of all the integers $R_i$ and that with every refreshment a
congruence system has to be solved, which can quite slow down the
rekeying process. Recall also that the server has to encrypt as many
times as the number of authorized users. In order to speed it up it is
commonly used jointly with HTA. However the length of rekeying
messages still depends on the users involved in each group.

As far as the Conditional Access Service introduced in \cite{cas} is
concerned, 
amid a good behavior regarding key storage, 
the high degree of the polynomials involved again generally forces a
partition of the users into
groups. 
Moreover the hash function used to create the interpolator polynomial
that is used to distribute the secret has to be changed with every
rekeying process, as mentioned above.

In our case, the rekeying process only requires a few simple
operations and is considerably faster with respect to all the
previously considered protocols. 



\section{Implementation of the proposed method}

The application was designed using three main computational objects, {\em the Key Sharing Framework object} (KSF), {\em a Server object} and
a {\em Client object}. The Server object is the hotspot in terms of computation due to the size of the matrix that it hosts (namely the vector space). The KSF manages clients and interfaces the GPU device, if present. The application was organized in the following stages:
\begin{itemize}
\item \emph{Vector space setup}: The KSF object creates the 2D matrix consisting of mutually orthogonal vectors.
\item \emph{Coder generation}: The Vector space can be reduced by column order into a 1-D vector (using the addition). This 1-D key is used by the Server to encode the content to be distributed.
\item \emph{User/Client login}: Prepares the necessary data structures to hold users claiming a key to decode content from the Server. Once the client is authorized to log in, the key to decode messages, is provided. 
\item \emph{Server initialization and startup}: The KSF framework permits the Server to accept requests from clients.
\end{itemize}

\begin{table*}[htpb]
  \caption{Execution of the protocol CPU-only threaded version (time in ms, vector and user \# in thousands, i.e. 5v x 5u stands for 5000v x 5000u)} 
  \centering 
  \small \begin{tabular}{|c | c c c | c c c|} 
    \hline\hline 
    \multicolumn{1}{|c}{} & \multicolumn{3}{|c|}{{\bfseries core i7 ee(12 hw threads)}} & \multicolumn{3}{|c|}{{\bfseries dualcore T9500 (2 hw threads)}}  \\
    {\bfseries stage} & 5v x 5u & 10v x 5u & 10v x 10u & 5v x 5u & 10v x 5u & 10v x 10u\\[0.5ex] \hline 
    Orthogonalization   & 114240    & 913866    & 913596    & 169909    & 1339753   & 1371083 \\ 
    Key Refreshment     & 30        & 197       & 198   & 66        & 231       & 235 \\
    Generator Coder     & 30        & 197       & 198   & 66            & 231       & 235 \\
    Server activation   & 0         & 1         & 0     & 2             & 1         & 3  \\
    Client setup      & 1         & 1         & 2   & 46            & 80        & 168 \\
    Broadcast   & 12        & 16        & 33    & 1 & 1 & 0 \\
    Client refresh & 1        & 1         & 1 & 45       & 300       & 556 \\
    \hline 
  \end{tabular}
  \label{table:th_i7_t9500} 
\end{table*}

\subsection{Results}
\begin{figure}
  \begin{center}
    \includegraphics[scale=0.3]{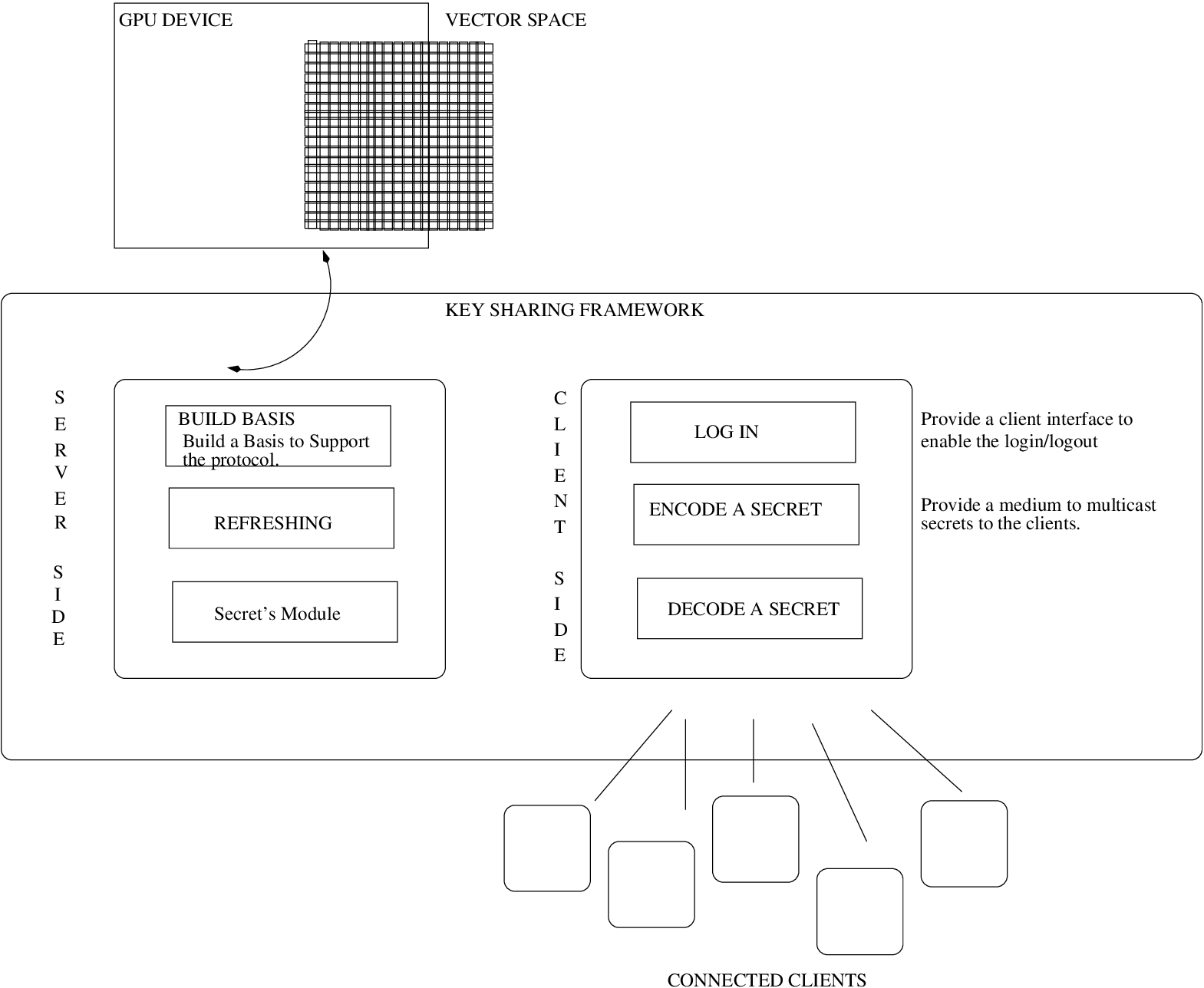}
    \caption{Cuda implementation schema}
    \label{fig:jcuda}
  \end{center}
\end{figure}

Jcuda (\cite{jcuda}), which is a Java binding for CUDA (Compute Unified Device Architecture), a set of development tools that allows programmers to use graphic cards for parallel computing-  was used to interface the GPU and impersonate it as a new computational entity to which we were able to send requests (cf. Figure \ref{fig:jcuda}).  To build and prepare the vector space Matrix in the GPU device, a kernel (code able to run on a GPU device) was written.
Table~\ref{table:th_i7_t9500} shows the execution scenarios and the timings in milliseconds for each protocol stage. Tests were executed on a Intel Core-i7 processor. The Server was run using vector spaces from 10 up to 10000. All the cases used 4 bytes for each vector component, so that the messages produced, using a vector space of dimension 10000, are less than 40 Kbytes. The orthogonalization stage is the most time consuming step, so setup is affected as it depends on the dimension of the vector space. To test the architectural benefits of the core i7, Table~\ref{table:th_i7_t9500}, the server was tested in a conventional processor, core 2 duo. Although this is a laptop processor's architecture, the trend reflected in Table~\ref{table:th_i7_t9500} follows the one observed in the core 2 duo case.

\begin{table}[h]
  \caption{Hardware acceleration of the orthogonalization process (in ms). The GPU used was a GForce GTX-460.} 
  \centering 
  \begin{tabular}{c c c c } 
    \hline\hline 
    & 5000vx5000u & 10000vx5000u & 10000vx10000u \\ [0.5ex] 
    \hline 
    GPU Orthog.     &    18878.9    &   39000.2 &  40183.4  \\ 
    GPU to CPU  &    89.3   &   130.3   &  221.2    \\
    \hline 
  \end{tabular}
  \label{table:jcudaGPU}
\end{table}
As Table~\ref{table:jcudaGPU} shows, the time spent in the Setup of the server, affected by the orthogonalization process, was reduced if a GPU was present.  Another issue was the client removal and the time to renew its key to make it available for new users. This time is reflected in the stage named \emph{Client setup}. As it can be seen, the time to refresh a client is not dependent on the dimension of the vector space. 

\section{Conclusions}

We have introduced a new protocol for managing keys in a centralized
secure multicast setting. This protocol is shown to be secure against
possible inner and/or outer attacks.  We also showed its advantages
with respect to other existing methods for key management in secure
multicast schemes, namely minimal requirements for computational
costs, key storage at both client and server sides, length and number
of rekeying messages per join and/or leave operation. Finally we
provided results showing that an efficient implementation is indeed
feasible.




\end{document}